Paper #90 – Estimation of poroelastic parameters from seismograms using Biot theory

*by Louis De Barros* [1,2] *and Michel Dietrich* [1,3]
(1) LGIT, CNRS, Univ. Joseph Fourier, 38041 Grenoble, France
(2) now at School of Geological Sciences, Univ. College Dublin, Belfield, Dublin 4, Ireland
(3) now at IFP, 1-4 Avenue de Bois Preau, 92852 Rueil Malmaison, France

**ABSTRACT**

We investigate the possibility to extract information contained in seismic waveforms propagating in fluid-filled porous media by developing and using a full waveform inversion procedure valid for layered structures.
To reach this objective, we first solve the forward problem by implementing the Biot theory in a reflectivity-type simulation program. We then study the sensitivity of the seismic response of stratified media to the poroelastic parameters. Our numerical tests indicate that the porosity and consolidation parameter are the most sensitive parameters in forward and inverse modeling, whereas the permeability has only a very limited influence on the seismic response.
Next, the analytical expressions of the sensitivity operators are introduced in a generalized least-square inversion algorithm based on an iterative modeling of the seismic waveforms. The application of this inversion procedure to synthetic data shows that the porosity as well as the fluid and solid parameters can be correctly reconstructed as long as the other parameters are well known. However, the strong seismic coupling between some of the model parameters makes it difficult to fully characterize the medium by a multi-parameter inversion scheme. One solution to circumvent this difficulty is to combine several model parameters according to rock physics laws to invert for composite parameters. Another possibility is to invert the seismic data for the perturbations of the medium properties, such as those resulting from a gas injection.

**INTRODUCTION**

Seismic wave imaging and inversion of the subsurface properties are usually carried out by assuming that the propagation media are acoustic or elastic media. These approximations are valid for the structural imaging of the subsurface, but are insufficient to deal with the reservoir engineering problems arising in hydrology, in the oil and gas industry and for the underground storage of carbon dioxide.
The Biot-Gassmann poroelasticity theory [1,2] opens up the possibility to characterize the porous media in greater detail, notably in terms of porosity, permeability and fluid properties, since the latter can be related to the seismic wave amplitudes and attenuation.

Classically, a two-step approach is employed to estimate the porous medium properties. The procedure uses the wave velocities which are first derived from the seismic data before being interpreted in terms of poroelastic parameters via a petrophysical model. Here, we develop an alternative approach in which the porous medium properties are 'directly' estimated from a full waveform inversion of the seismic data.

Several ingredients are required to solve this problem. We first need a forward modeling code to compute the point source response of a fluid-saturated, layered poroelastic medium. Second, we must know the sensitivity operators of the seismic waves with respect to the properties of the porous medium. Finally, we have to define and implement an optimization technique to reconstruct the medium properties by means of an iterative inversion scheme.

## WAVE PROPAGATION IN POROUS MEDIA

### Poroelastic theory

The equations of poroelasticity, initially developed by Biot [1] have been rewritten [3] in the form :

$$[ (K_U + G/3)\nabla\nabla + (G\nabla^2 + \omega^2 \rho) \mathbf{I} ] \cdot \mathbf{u} + [ C\nabla\nabla + \omega^2 \rho_f \mathbf{I} ] \cdot \mathbf{w} = 0 ,$$
$$[ C\nabla\nabla + \omega^2 \rho_f \mathbf{I} ] \cdot \mathbf{u} + [ M\nabla\nabla + i \omega \eta \mathbf{I} / k(\omega)] \cdot \mathbf{w} = 0 . \qquad (1)$$

We refer the reader to the work of Pride [3] for the definition of the variables involved in eq. (1) and for the Gassmann decomposition of the undrained parameters [2]. The properties considered to describe the poroelastic medium are [4] 1) the porosity $\varphi$, 2) the mineral bulk modulus $K_s$, 3) the mineral density $\rho_s$, 4) the mineral shear modulus $G_s$, 5) the consolidation factor $cs$ , 6) the fluid bulk modulus $K_f$, 7) the fluid density $\rho_f$ , 8) the permeability $k_0$ and 9) the fluid viscosity $\eta$. With this decomposition, we can easily reanrange them to obtain new parameters as the saturation rate Sr (ratio of the volume occupied by the most viscous fluid normalized by the pore volume) or the mineral rate Ts (volume occupied by a mineral normalized by the total mineral volume).

### Forward problem

The forward problem, i.e., the computation of synthetic seismograms from the poroelastic theory, has been tackled with several techniques. We use here the generalized reflection and transmission matrix method of Kennett [5] to solve this problem in the frequency-wavenumber domain for horizontally layered media. The synthetic seismograms are finally obtained in the time-distance domain by using the discrete wavenumber integration technique of Bouchon [6].

This approach was first used by Garambois and Dietrich [7] who considered seismic and electromagnetic waves propagating in porous media. Our modeling code is a simplified version of this simulation program limited to seismic waves only. This computation method accurately accounts for all multiple reflections and mode conversions, and includes surface waves and near-field effects.

## SEISMIC WAVE SENSITIVITY

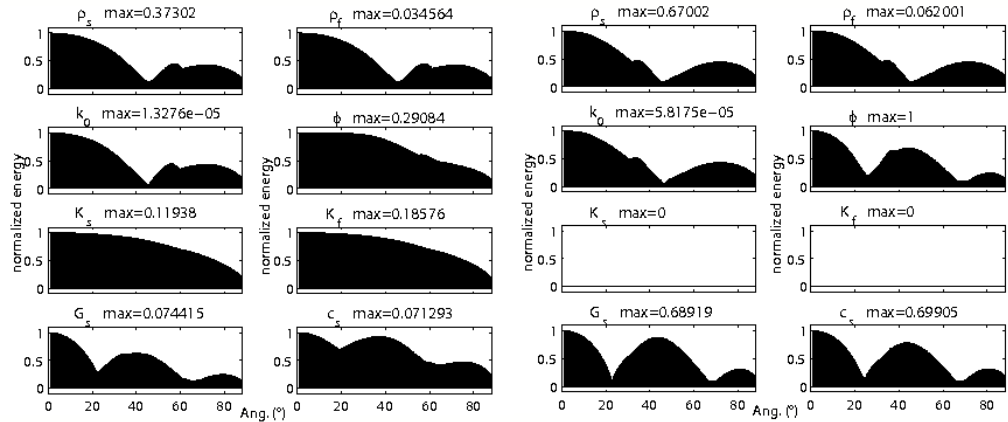

Figure 1. Energy of reflected waves as a function of incidence angle (AVA response) for perturbations in parameters $\rho_s$, $\rho_f$, $k_0$, $\varphi$, $K_s$, $K_f$, $G_s$ and cs. The eight left (resp. right) panels are for PP (resp. SS) reflections. The maximum amplitudes of the curves are given above each panel.

The sensitivity of the seismic waveforms to small changes of the model parameters is investigated by computing the first-order derivatives of the seismic displacements with respect to the relevant poroelastic parameters described in the previous section. These operators, which are often referred to as the Fréchet derivatives, can be expressed via semi-analytical formulas by using the Born approximation [4].

Figure 1 presents the Amplitude Versus Angle (AVA) curves for the *PP* (left) and *SS* (right) reflected waves due to a small and localized perturbation of a model parameter.

We note that the reflected waves are especially sensitive to the mineral density $\rho_s$, porosity $\varphi$, shear modulus $G_s$ and consolidation parameter *cs*. If the mineral properties ($G_s$, $K_s$ and $\rho_s$) are fixed, the porosity $\varphi$ and consolidation factor *cs* are the most sensitive parameters and therefore the main parameters to consider in an inversion procedure. On the other hand, the viscosity $\eta$ and permeability $k_0$ have only a weak influence on the wave amplitudes. For the *P*-waves, the fluid modulus $K_f$ has a stronger influence than the solid modulus $K_s$ if the medium is poorly consolidated (as in Figure 1). The inverse is true for a consolidated medium. We note in Figure 1 that for some parameters, the model perturbations lead to similar modifications of the seismic response. For example, perturbations in densities and permeability show identical AVA responses. The same is true for the bulk moduli. Parameters that exhibit this behavior cannot be properly reconstructed in the same inversion procedure.

## FULL WAVEFORM INVERSION

### Inversion algorithm

Our method to determine the intrinsic properties of porous media is based on a full waveform iterative inversion procedure. It is carried out with a gradient technique to infer an optimum model which minimizes a misfit function. The latter is defined by a sample-to-sample comparison of the observed data $\mathbf{d_{obs}}$ with a synthetic wavefield $\mathbf{d} = f(\mathbf{m})$ in the time-space domain, and by an equivalent term describing the deviations of the current model $\mathbf{m}$ with respect to an a priori model $\mathbf{m_0}$, i.e.,

$$S(\mathbf{m}) = \tfrac{1}{2} \{ \| \mathbf{d} - \mathbf{d_{obs}} \|_D + \| \mathbf{m} - \mathbf{m_0} \|_M \}, \qquad (2)$$

where the L2-norms $\| \cdot \|_D$ and $\| \cdot \|_M$ are defined in terms of a data covariance matrix $\mathbf{C_D}$ and an a priori model covariance matrix $\mathbf{C_M}$ [8]. The model $\mathbf{m}$ contain the description of one or several parameters in layers which thickness is defined by the peak content of the data [9].

The model is updated using a quasi-Newton algorithm, which involves the Fréchet derivatives obtained earlier. As this problem is strongly non-linear, several iterations are necessary to converge toward an optimum model $\underline{\mathbf{m}}$, i.e, a model whose response $\underline{\mathbf{d}}$ satisfactorily fits the observed data $\mathbf{d_{obs}}$.

**Numerical Results**

In the following examples, we consider layered media excited by a vertical point force with a perfectly known source time function. The computed synthetic seismograms are vertical displacements generated at various source-receiver offsets. Source and receivers are located at the free surface. Direct waves and surface waves are not included in our simulations to simplify the datasets and avoid any pre-processing to remove these contributions.

We first assume that only one model parameter distribution is unknown, the other parameters being perfectly known and fixed. In this case, our inversion procedure yields very satisfactory results: the computed seismograms obtained at the end of the inversion process are very close to the synthetic data to be inverted, and the reconstructed model perfectly fits the true model. The only parameter which is difficult to reconstruct is the permeability because of its weak influence on seismic waves.

When trying to invert for several parameters at the same time, we end up with seismograms that correctly fit the data, but the algorithm fails to reconstruct the correct model parameter distributions, because of the coupling between some of the parameters. Therefore, we have to find strategies to circumvent this difficulty.

**Inversion Strategy**

NEW MODEL PROPERTIES

One way to get round the difficulty of multi-parameter reconstruction is to use external information and to combine parameters that are physically connected. For example, when porous media are filled with biphasic fluids (e.g., water and air), the problem becomes tractable if we know the nature of the fluids. We can then invert the seismic data to determine the saturation rate *Sr* . We can similarly invert for the rate *Ts* of a mineral when the medium considered is constituted by two minerals. Figure 2 shows inversion results for this parameter when the constitutive minerals are silica and mica. The computations are carried out to determine the rate of silica,the others parameters are perfectly known. We note that, in this case, the model as well as the data is very well reconstructed.

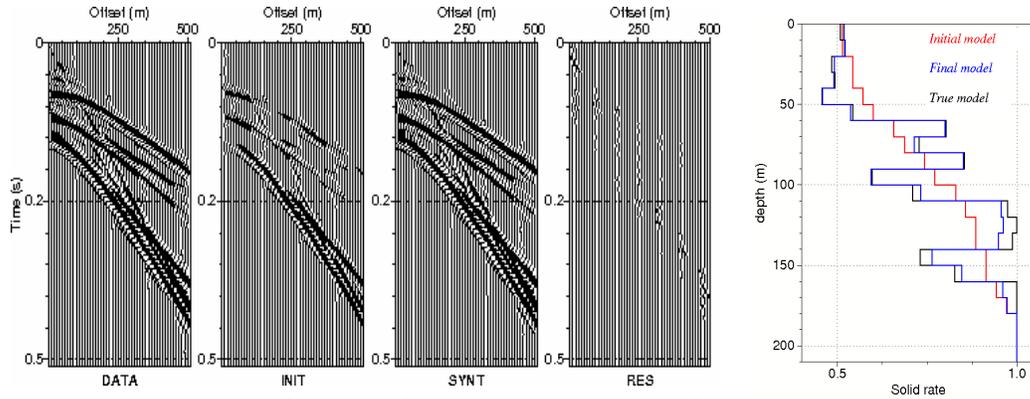

Figure 2. Inversion of the rate of silica. The panel on the left shows the initial model which is also the a priori model (red), the true model (black) and the final model obtained (blue). The seismic sections on the right display, from left to right, the synthetic data to invert (DATA), the seismograms corresponding to the initial model (INIT), the seismograms obtained at the end of the inversion process (SYNT), and the residual data (RES).

If the medium is made up of several lithologic facies (sand and shale for example) which are known beforehand, we can invert for the percentage of each lithology. In this case, the maximum number of unknowns to fully characterize the medium is reduced to two (one for the fluid, one for the solid).

DIFFERENTIAL INVERSION

Time-lapse monitoring surveys are designed to investigate fluid substitutions in reservoir rocks. The approach followed here consists in performing a preliminary inversion of synthetic data corresponding to the reference medium prior to the fluid change. As mentioned previously, multi-parameter inversions yield poor results in terms of model properties, but the wave forms are reasonably well estimated. The model thus obtained may therefore be used as the starting model to follow the fluid substitution. We carried out a second inversion for the saturation rate of the fluid of interest, after a change of the fluid content. As shown in Figure 3, this inversion yields good results: the fluid variations are perfectly localized even though the properties of the solid medium are poorly determined.

**CONCLUSION**

The main objective of this work was to develop a methodology to estimate the intrinsic properties of porous media from a full waveform inversion of seismic data. We have successively developed the various ingredients required by this approach. First, we have solved the forward problem by extending the reflectivity method to poroelastic media. Second, we have derived sensitivity operators in the form of semi-analytical formulas involving the Green's functions of the unperturbed media. Third, we have implemented these tools in an iterative least-square (quasi-Newton) inversion algorithm.

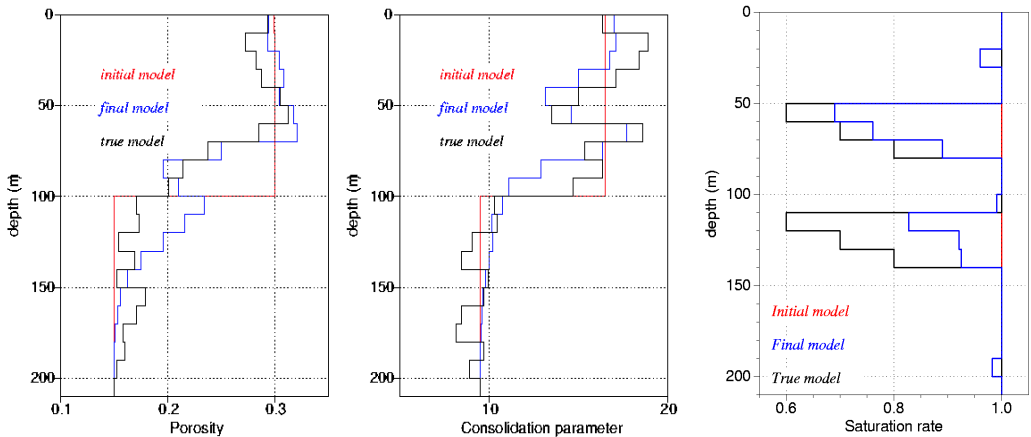

Figure 3. Starting model (red), true model (black) and final model (blue) for the porosity (left panel) and consolidation factor (middle panel) in a five-parameter inversion performed to determine the reference medium in a time-lapse survey. The panel on the right shows the saturation rate as a function of depth obtained after a fluid change by using the model properties estimated in the first inversion.

Our inversion algorithm proved efficient to reconstruct the variations of a single parameter when all other parameters are perfectly known. However, strong coupling between model parameters did not allow us to simultaneously invert for several parameters. To circumvent this problem, external information must be used and alternative parameters such as the lithology or the saturation rate must be considered. Finally, we have shown that for time-lapse surveys, a sequence of two inversions, before and after fluid substitution, is capable of properly estimating the fluid content in the subsurface even though the properties of the reservoir rocks are imperfectly described.